%% file: ENIEF2013-3496_paper.tex
\documentclass[oneside,a4paper,english,links]{amca}

\usepackage[utf8]{inputenc}
\usepackage{ae,aecompl}
\usepackage{graphicx}
\usepackage{subfigure}
\usepackage{amsmath,amsfonts}
\usepackage{commath}
\usepackage{dsfont}
\usepackage{cases}
\usepackage{bm}
\usepackage{booktabs}

\title{Identification of parameters in the torsional dynamics of a drilling process through Bayesian statistics}

\author[a,b]{Mario Germ\'{a}n Sandoval}
\author[a]{Americo Cunha Jr}
\author[a]{Rubens Sampaio}
\affil[a]{Department of Mechanical Engineering, PUC--Rio\newline
Rua Marqu\^{e}s de S\~{a}o Vicente, 225, G\'{a}vea, Rio de Janeiro - RJ, Brazil.\newline
americo.cunhajr@gmail.com ~~ rsampaio@puc-rio.br}
\affil[b]{Departamento de F\'isica, Universidad Nacional del Sur\newline
Avenida Alem 1253 - Bah\'ia Blanca (CP 8000) - Pcia. de Buenos Aires - Argentina.\newline
sandomario@gmail.com}

\input{macros}

\begin{document}
\vspace{3cm}

\maketitle


\begin{abstract}

This work presents the estimation of the parameters of an experimental setup,
which is modeled as a system with three degrees of freedom, composed by a 
shaft, two rotors, and a DC motor, that emulates a drilling process.
A Bayesian technique is used in the estimation process, to take into account 
the uncertainties and variabilities intrinsic to the measurement taken, which are 
modeled as a noise of Gaussian nature. With this procedure it is expected to check 
the reliability of the nominal values of the physical parameters of the test rig. 
An estimation process assuming that nine parameters of the experimental 
apparatus are unknown is conducted, and the results show that for some 
quantities the relative deviation with respect to the nominal values is very high. 
This deviation evidentiates a strong deficiency in the mathematical model used to 
describe the dynamic behavior of the experimental apparatus.

\end{abstract}

\section{INTRODUCTION}

A drillstring is a device, used to drill oil wells, whose three-dimensional 
dynamics behavior is very complex, and subject to three mechanisms of vibration: 
longitudinal, transverse, and torsional \citep{spanos2003p85}. The torsional mechanism 
is responsible for causing a type of dynamics called stick-slip. In this mode, 
the rotation of the column is blocked due to dry friction with the borehole, stick, and then suddenly 
released, slip, to be afterwards blocked again. So, due to the constant rate of rotation imposed at the 
top of the column, the column is severely twisted. This may cause fatigue damage, 
or even the rupture of the structure.

To characterize the stick-slip mode and, consequently, to develop 
mechanisms to avoid it, the works of \cite{CAYRES.B1} and \cite{cayres2013} present 
numerical studies which investigate, by means of a sensitivity analysis, the influence 
of the friction models and the caracterization of the structure parameters during the 
occurrence of the stick-slip. Experimental data obtained from a test apparatus, in laboratory 
scale, are used to validate the mathematical model developed.

One of the challenges associated to the use of an experimental setup to validate a mathematical 
model is the correct characterization of the parameters that are used to describe the behavior 
of the experimental setup physical system. Commonly, a physical phenomenon is not described 
perfectly by a mathematical model, which may be due to an inability of the model itself, or also,
due to inaccuracies of the measuring instruments, ambient noise, human errors, 
and several others factors. In this way, the measures that one has for estimate the 
parameters of a physical system are subjected to variabilities~\citep{vanderheijden2004}. 
Fortunately, the measurements erros can be circumvented with the use of stochastic techniques 
of estimation, in particular, the Bayesian inference, which updates the estimative of the parameters 
for each new evidence collected about them~\citep{William-2007}.


In order to characterize the experimental setup presented in 
\cite{CAYRES.B1} and \cite{cayres2013}, that emulates a drilling process, this paper 
presents a study in which the physical parameters associated with the test rig are 
estimated through the Bayesian technique. The main objective that one wants to achieve
with the parametric estimation is to check the reliability of the nominal values of the
physical parameters of interest.

This paper is organized as follows. In section~\ref{experim_setup}
the experimental apparatus is described. The mathematical modeling
of the physical system appears in section~\ref{mat_modeling}. In section~\ref{param_estimation}, 
the Bayesian estimation technique is briefly described. Then, in section~\ref{results_discussion},
the estimation results for the experimental setup are presented and discussed. 
Finally, in section~\ref{concl_remaks}, 
the main conclusions are emphasized and some directions for future work outlined.

\section{EXPERIMENTAL SETUP}
\label{experim_setup}

\begin{figure}[h]
  \centering
  \includegraphics[scale=0.5]{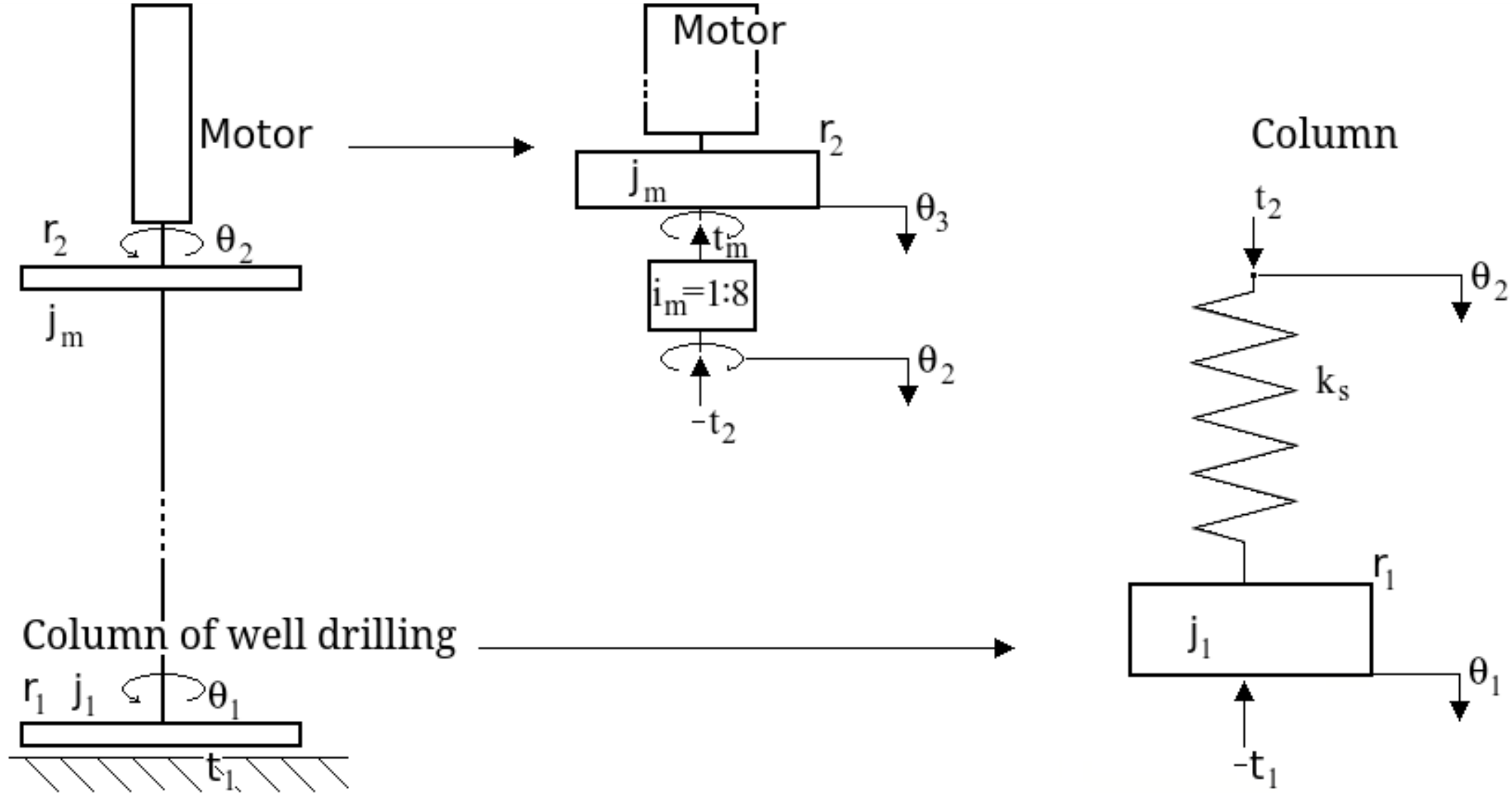} 
  \caption{Schematic representation of the experimental setup used to emulates a drilling process.}
  \label{fig_experimental_setup}
 \end{figure}

The experimental apparatus used in the works of
\cite{CAYRES.B1} and \cite{cayres2013}, which is intended to emulate 
a drilling process, basically consists of a large flexible shaft, one DC motor 
and two rotors, as can be seen in the schematic representation of
the Figure~\ref{fig_experimental_setup}. 

In this experimental setup, the flexible shaft has a very low flexural stiffness, 
so that it emulates the behavior of a drillstring, which is a very slender structure 
due to the high value of the ratio $ l_{s}/D_{s}$, between the column length, $l_{s}$,
and the diameter of the column cross section, $D_{s}$. 

\pagebreak
The DC motor has the role 
of imposing a constant speed on top of the column, so that it can penetrate the rock. 
The rotor $r_2$ is located between 
the DC motor and the shaft, and concentrates the inertia of the motor, which 
is denoted by $j_m$. On the other extreme of the shaft one can find the rotor 
$r_1$, which has an inertia $j_1$.

The modeling of the experimental setup assumes that it has three degrees of freedom, 
being the first degree of freedom (DoF) the angle of rotation of the shaft
around the rotor $r_1$, which is denoted by $\theta_1$, the second DoF
is the angle $\theta_2$, which measures the shaft rotation with respect 
to the rotor $r_2$, and the last DoF is the the electrical charge in the motor,
denoted by $q$. The other physical parameters of this mechanical system, 
as well as their nominal values, are presented in Table~\ref{tab_model_parameters}.

For further details about this experimental apparatus and the model that is used
to describe the experiments, the reader is encouraged to see the work of \cite{cayres2013}.

\begin{table}[h]
	\centering
    \begin{tabular}{c l r l}
    \toprule
     \multicolumn{4}{c}{\bf Column}\\
	\midrule
	parameters & description & value & unit  \\
	\midrule
\noalign{\medskip}  $  j_{1}   $   &  moment of inertia &    $28.3 \times 10^{-3} $  &  $ kg\,m^2  $    \\
\noalign{\medskip}  $  l_{s}   $   &  length of the column &    $     2.4      $  &  $    m     $    \\
\noalign{\medskip}  $  D_{s}   $   & diameter of the column &    $    3     $  &  $    mm     $    \\
\noalign{\medskip}  $ m_{r_1}  $   & mass of the rotor 1            &    $    6.4     $  &  $    kg    $    \\
\noalign{\medskip}  $    k_s   $   &   torsional stiffness of the column &    $  3.0 \times 10^{-1}    $  &  $  Nm/rad  $    \\
\noalign{\medskip}  $ r_{r_1}  $   & radius of the rotor 1           &    $    188     $  &  $    mm     $    \\
\noalign{\medskip}  $ t_{1}    $   &   friction torque in the rotor 1      &   $0$   &  $    N m     $    \\
& & &\\
 \multicolumn{4}{c}{\bf Motor}\\
	\midrule
	parameters & description & value & unit  \\
	\midrule
\noalign{\medskip}  $  j_{m}   $   & moment of inertia &   $ 4.0 \times 10^{-4} $  &  $ kg\,m^2  $    \\
\noalign{\medskip}  $  l_{m}   $   & internal inductance of the motor &   $1.1 \times 10^{-3}  $  &  $    H     $    \\
\noalign{\medskip}  $  r_{m}   $   & internal resistance of the motor &   $3.3 \times 10^{-1}     $  &  $ \Omega   $    \\
\noalign{\medskip}  $   k_T    $   & constant of torque          &   $1.2 \times 10^{-1}$  &  $  Nm/A    $    \\
\noalign{\medskip}  $   k_e    $   & constant of voltage &   $601.6 \times 10^{-4}$  &  $V/(rad/s) $    \\              
\noalign{\medskip}  $   t_f    $   & torque due to motor internal friction &   $     1.0 \times 10^{-1}      $  &  $    Nm    $    \\
\noalign{\medskip}  $   c_m    $   & internal damping of the motor &   $ 18.0 \times 10^{-5} $  &  $Nm/(rad/s)$    \\              
\noalign{\medskip}  $   i_m    $   & gear reduction factor &   $     1/8       $  &  $   8:1    $    \\
\noalign{\medskip}  $    v     $   &  voltage &   $      8       $  &  $    V     $    \\
\bottomrule
    \end{tabular}
	\caption{Numerical values for the physical parameters of the experimental setup.}
	\label{tab_model_parameters}
\end{table}


\section{MATHEMATICAL MODELING}
\label{mat_modeling}

The following lines reproduce the mathematical modeling,
developed by \cite{cayres2013} and \cite{CAYRES.B1}, of the 
experimental setup described in the last section.

Starting the equationing by the motor, one has that
the sum of the electric potentials in the motor is equal to the 
external potencial applied on it, so that

\begin{equation}
l_m \ddot{q}(t) + r_m \dot{q} (t) + k_e \dot{\theta}_3 (t)  = v,
\label{eq:bancada.motor.1}
\end{equation}

\noindent
where $l_m$ is the internal inductance of the motor, 
$r_m$ is the internal resistance of the motor,
$k_e$ is a constant of voltage, and
$\theta_3$ is the angular rotation of the motor axis.
Note that the upper dot denotes differentiation with 
respect to time.
 
The balance of the torques acting on the motor provides the following equation
 
 \begin{equation}
  j_m \ddot{\theta}_3 (t) + c_m \dot{\theta}_3 (t) - k_T \dot{q} (t) + t_f  = - t_m,
 \label{eq:bancada.motor.2}
 \end{equation}
 
\noindent
where $c_m$ is the internal damping of the motor,
$k_T$ is a constant of torque,
$t_f$ is the torque due to motor internal friction,
and $t_m$ is the torque on the motor due to the coupling
between the motor and the gearbox.

\pagebreak
Due to the reduction factor in the gearbox, there is a relationship between 
the angles of rotation $\theta_2$ and $\theta_3$, which is given by

\begin{equation}
 i_m  \theta_3 = \theta_2,
 \label{eq:bancada.reduc.ang}
\end{equation}

\noindent
being $i_m$ the gear reduction factor. This reduction factor also provides 
a coupling between $t_m$ and the torque on the second DoF, $t_2$,
which is expressed by

\begin{equation}
 t_m = i_m t_2.
 \label{eq:bancada.reduc.torq}
\end{equation}

Moreover, the balance of the torques which are acting on the rotor $r_2$ provides 

\begin{equation}
  - k_s \left( \theta_1(t) - \theta_2 (t) \right) = t_2,
 \label{eq:bancada.coluna.r2}
\end{equation}

\noindent
being $k_s$ is the torsional stiffness of the shaft.

Likewise, the balance of the
torques acting on the rotor $r_1$ provides the following differential equation

\begin{equation}
 j_{1} \ddot{\theta}_1 (t) + k_s \left( \theta_1(t) - \theta_2(t) \right) = -t_1.
 \label{eq:bancada.coluna.r1}
\end{equation}

Combining the Eqs.(\ref{eq:bancada.motor.2}) to (\ref{eq:bancada.coluna.r2}),
one can write

\begin{equation}
   j_{m} \ddot{\theta}_2(t) - i_{m}^{2} k_s \left( \theta_1(t) - \theta_2(t) \right)
   - i_m k_T \dot{q}(t) + c_m \dot{\theta}_2 = - i_m t_f.
  \label{eq:bancada.3}
 \end{equation}

Thus, through a rearrangement of the Eqs.(\ref{eq:bancada.motor.1}), 
(\ref{eq:bancada.coluna.r1}), and (\ref{eq:bancada.3}), it is possible represent 
the dynamics of the experimental setup by the following state space equation

\begin{eqnarray}
  \dot{\vec{y}}(t) = \mat{A} \vec{y}(t) + \vec{F}(t),
  \label{eq_state_space}
\end{eqnarray}

\noindent
where the state vector $\vec{y}(t)$, the control vector $\vec{F}(t)$, and 
the system matrix $\mat{A}$ are, respectively, defined as

\begin{equation}
	\vec{y}(t) = 
 \begin{pmatrix}
           \theta_1(t)   \\
           \theta_2(t)   \\
                q(t)     \\
      \dot{\theta}_1(t)  \\
      \dot{\theta}_2(t)  \\
      \dot{q}(t)
 \end{pmatrix},
~~
\vec{F}(t) = 
 \begin{pmatrix}
	           0          \\
	           0          \\ 
	           0          \\
	       -t_1/j_1       \\
	       -i_m t_f/j_m   \\
	         v/l_m 
 \end{pmatrix},
 \end{equation}

\noindent
and 
 
 \begin{equation}
\mat{A} = 
 \begin{pmatrix}
        0    &           0    &   0   &   1   &     0       &     0          \\
        0    &           0    &   0   &   0   &     1       &     0          \\
        0    &           0    &   0   &   0   &     0       &     1          \\   
    -k_s/j_1 &        k_s/j_1 &   0   &   0   &     0       &     0          \\
  {i_m^2}k_s/j_1 &-{i_m^2}k_s/j_m &   0   &   0   & -c_m/j_m    &  i_m k_T/j_m   \\
        0    &           0    &   0   &   0   &-i_n k_e/l_m & -r_m/l_m
 \end{pmatrix}.
\end{equation}

Later in this work will be necessary to integrate the initial value problem
defined by the ODE of the Eq.(\ref{eq_state_space}), and an appropriate
initial condition, assumed as the zero vector for all the simulations
reported in this work. This integration is carried out numerically,
using the routine \textsc{ode23t} of MATLAB, that uses a trapezoidal rule 
with ``free'' interpolant to construct the time response of the dynamical 
system under analysis. The choice of this ODE solver was motivated
by the presence of a moderately stiffness in the system of equations
to which one wish to solve.

\section{PARAMENTERS IDENTIFICATION}
\label{param_estimation}

The Bayesian statistical inference theory provides the link between the results of 
the experimental observations with the theoretical predictions. This can be 
tapped for parametric estimation techniques. Thus, given the known 
distribution of the observations conditioned on the parameters $p(\vec{y}|\vec{x})$ 
(forward PDF), it is desired to find a probability distribution of the parameters 
influenced by the observations $p(\vec{x}|\vec{y})$ (posterior PDF),~\citep{William-2007}.

Thus,

\begin{equation}
 p(\vec{x}|\vec{y}) = \frac{1}{p(\vec{y})} \, p(\vec{y}|\vec{x}) \, p(\vec{x}) \;,
 \label{Eq:T.Bayes}
\end{equation}
where $\vec{x}$ is the parameters vector and $\vec{y}$ is the sample vector.

The Eq.(\ref{Eq:T.Bayes}) is the Bayes theorem, and states that the posterior PDF 
can be obtained from the forward PDF, determined by the experiment (or simulation).
The a priori PDF, $p(\vec{x})$, is calculated based on what is known (or believed to be known) 
about the parameters before observations. The ratio $1/p(\vec{y})$ is a normalization constant.

To obtain the desired parametric estimation through Bayesian inference, it is necessary 
to seek the set of parameter that maximize the Eq.(\ref{Eq:T.Bayes}). For this purpose, the 
natural logarithm is applied to the posterior PDF, considering Bayes law and under 
the assumption that the values of the parameters before the observation do not 
change i.e., $p(\vec{x})$ is constant. In this way one obtains that

\begin{equation}
  \ln{\left(p(\vec{x}|\vec{y})\right)} \propto \ln{\left(p(\vec{y}|\vec{x})\right)},
 \label{Eq:log.T.Bayes}
\end{equation}

\noindent
that is, the posterior PDF is proportional to the forward PDF.
Therefore, to maximize the left hand side of (\ref{Eq:log.T.Bayes})
is similar to maximize its right hand side.

In general, the measurements of an experiment have some type of dispersion, 
associated with each observation, and this dispersion can be modeled as an 
error of random nature, which will be called $\vec{n}$. This error is added to the 
value given by the physical theory $\hat{\vec{y}}$, which is considered free of error. 
Thus, one has

\begin{equation}
 \vec{y} = \hat{\vec{y}} + \vec{n}.
 \label{Eq:amostras}
\end{equation}

It is assumed that the first two moments of $\vec{n}$ are known (the average value $\mu_n$ 
and the covariance matrix $\mat{{\Gamma}_n}$) and that its distribution is a Gaussian. 
Also, it is supposed that the measurements are independent and identically distributed 
(i.i.d.) events, so that the covariance matrix is $\mat{{\Gamma}_n} = \sigma_n^2 \mat{I}$, 
where $\mat{I}$ is the identity matrix and $\sigma_n$ is the Gaussian standard deviation.

Therefore, the misfit function is defined as

\begin{equation}
 \varepsilon(\vec{x} \,;\; \vec{y}) = \frac{1}{{\sigma}_n^2}(\vec{y} - \hat{\vec{y}})^T(\vec{y} - \hat{\vec{y}})
                                = \frac{1}{{\sigma}_n^2} {\rVert{ \vec{y} - \hat{\vec{y}} }\lVert}^2,
 \label{Eq:min.cuad}
\end{equation}

\noindent
and must be minimized in order to estimate the parameters. This procedure is called
least-squares method.

The procedure followed in order to estimate the parameters of the experimental
apparatus consists of collecting the experimental data, construct the misfit function,
Eq.(\ref{Eq:min.cuad}), and then minimize it. The minimum of the function 
given by the Eq.(\ref{Eq:min.cuad}) is searched with the aid of the 
function \textsc{fminsearch} of MATLAB, which used M-simplex method 
\citep{Nelder-Mead-1965}.

For further information, the reader can see 
\cite{vanderheijden2004}.


\pagebreak
\section{RESULTS AND DISCUSSION}
\label{results_discussion}

\subsection{Verification of the technique effectiveness}

First, to verify the effectiveness of the method, only two parameters are
estimated, namely the constant voltage $k_e$ and the 
theoretical internal damping of the motor $c_m$.
Thus, the parameter vector is written as

\begin{equation}
  \vec{x} = 
  \begin{pmatrix}
                 c_m  \\ 
                 k_e
  \end{pmatrix}.
  \label{eq:bancada.prmtrs.minquad}
\end{equation}

\noindent
The other parameters values are are assumed to known,
and are obtained from the Table~\ref{tab_model_parameters}. 

Given that, in the laboratory, it is only possible to measure the values of 
the rotations, $\theta_1$ and $\theta_2$, and their respective time derivatives, 
$\dot{\theta}_1$ and $\dot{\theta}_2$ , a vector of samples is

\begin{equation}
   \vec{y}   =   {\begin{pmatrix} \theta_1 & \theta_2 & \dot{\theta}_1 & \dot{\theta}_2 \end{pmatrix}}^T  
              =  \hat{\vec{y}} + \vec{n},
   \label{eq:bancada.minquad.Z}
\end{equation}

\noindent
where $\hat{\vec{y}}$ is given by the numerical solution of the Eq.(\ref{eq_state_space}), and
the noise $\vec{n}$, as previously stated, is assumed to be i.i.d. and Gaussian.

Thus, the misfit function, according to Eq.(\ref{Eq:min.cuad}), is given by

\begin{equation}
   \varepsilon(\vec{x};\vec{y}) = \left( \frac{1}{\sigma_n^2} \right)   
             \sum_{i=1}^{2} \left[ \left( \bm{\theta}_i - \hat{\bm{\theta}}_i(\vec{x}) \right)^2 +
                                   \left( \dot{\bm{\theta}}_i - \hat{\dot{\bm{\theta}}}_i{(\vec{x})} \right)^2 \right]
   \label{eq:bancada.minquad.desfas}
\end{equation}

\noindent
with $i = 1,2$. In this effectiveness verification test, the $\hat{\bm{\theta}}_i(\vec{x})$ and 
$\hat{\dot{\bm{\theta}}}_i{(\vec{x})}$ are given by the mathematical model, while $\bm{\theta}_i$ and
$\dot{\bm{\theta}}_i$ are obtained from $\hat{\bm{\theta}}_i(\vec{x})$ and 
$\hat{\dot{\bm{\theta}}}_i{(\vec{x})}$ by adding a noise.

Estimates of the two unknown parameters, for different values of the standard deviation
$\sigma_n$, using as initial guess $c_m = 0.001$ and $k_e = 0.01$, and the corresponding 
deviation, computed with respect to the nominal values of the 
Table~\ref{tab_model_parameters} can be seen in Table~\ref{tab_estimation_2param}.

The results reported in Table~\ref{tab_estimation_2param} show a good agreement between 
the values estimated by the Bayesian technique and the reference values shown in
Table~\ref{tab_estimation_2param}. Relative deviations less than 1\% can be seen
in the case of $c_m$, and negligible values are observed for $k_e$. These results provide
an indication that the available nominal values for $c_m$ and $k_e$ are realistic.

\begin{table}[h]
  \centering
  \begin{tabular}{lcccc}
	\toprule
	& \multicolumn{2}{c}{$c_m$} & \multicolumn{2}{c}{$k_e$}\\
	\cmidrule(r){2-3} \cmidrule(r){4-5}
	$\sigma_n$	&	estimation & relative deviation & estimation & relative deviation\\
	\midrule
	 0.001 & $18.9 \times 10^{-5}$ & $0.77 \, \%$ & $601.6 \times 10^{-4}$ & $0.0064 \, \%$	\\
	   0.01 & $19.1 \times 10^{-5}$ & $0.77 \, \%$ & $601.6 \times 10^{-4}$ & $0.0068 \, \%$	\\
	     0.1 & $19.0 \times 10^{-5}$ & $0.60 \, \%$ & $601.6 \times 10^{-4}$ & $0.0051 \, \%$	\\
	     1.0 & $18.9 \times 10^{-5}$ & $0.82 \, \%$ & $601.9 \times 10^{-4}$ & $0.0061 \, \%$	\\
\bottomrule
  \end{tabular} 
  \caption{Parameters of the experimental setup estimated
  for different values of $\sigma_n$,  using as initial guess $c_m = 0.001$ and $k_e = 0.01$.}
  \label{tab_estimation_2param}
 \end{table}

\subsection{Estimation of the experimental apparatus parameters}

In this section are presented the results of the estimation of all the nine physical parameters 
that are associated to the experimental setup. In this way, the parameter vector 
is given as follows

\begin{equation}
    \vec{x} =
  {\begin{pmatrix}
        j_m  &  c_m  &  k_e  &  r_m  &  k_t  &  l_m  &  k_s  &  j_1  &  t_f
  \end{pmatrix}}^T.
  \label{eq:bancada.prmtrs.minquad.exp}
\end{equation}

For this case the expression of the misfit function is similar to the
one presented in Eq.(\ref{eq:bancada.minquad.desfas}), but now
the sample vectors are no longer calculated based on the 
mathematical model, but are obtained experimentally.

The first thing to notice is that when one tries to estimate all the nine 
parameters, simultaneously, using the Bayesian approach used in the 
estimation procedure reported in the last section, convergence problems 
in the minimization procedure are observed. 

To circumvent these problems of convergence, a heuristic procesudre 
to minimize the misfit function is proposed. 
In this heuristic algorithm, first it is assumed that the parameters 
$\vec{x}_1 = (t_f\;j_m)^T$ are unknown, and that the other parameters
are given by the values of the Table~\ref{tab_model_parameters}. 
Then, to obtain estimates for $t_f$ and $j_m$, the least-squares method 
is executed for 10 iterations, using an arbitrary initial guess.

In the next step of the heuristic algorithm, the unknown vector
is assumed to be $\vec{x}_2 = (j_m\;c_m)^T$. The value of
the parameter $t_f$ and the initial guess for $j_m$ are given by 
the estimation of the last step. Again the least-squares method 
is executed for 10 iterations, and an estimation for $j_m$ and $c_m$
is obtained.

In what follows, the procedure above continues using the following order to the 
vector of unknowns: $\vec{x}_3 = (c_m\;k_e)^T$, $\vec{x}_4 = (k_e\;k_t)^T$, 
$\vec{x}_5 = (k_t\;r_m)^T$, $\vec{x}_6 = (r_m\;l_m)^T$, $\vec{x}_7 = (l_m\;k_s)^T$, 
$\vec{x}_8 = (k_s\;j_1)^T$,  $\vec{x}_9 = (j_1\;t_f)^T$.

\begin{figure}[h]
  \centering
  \includegraphics[scale=0.5]{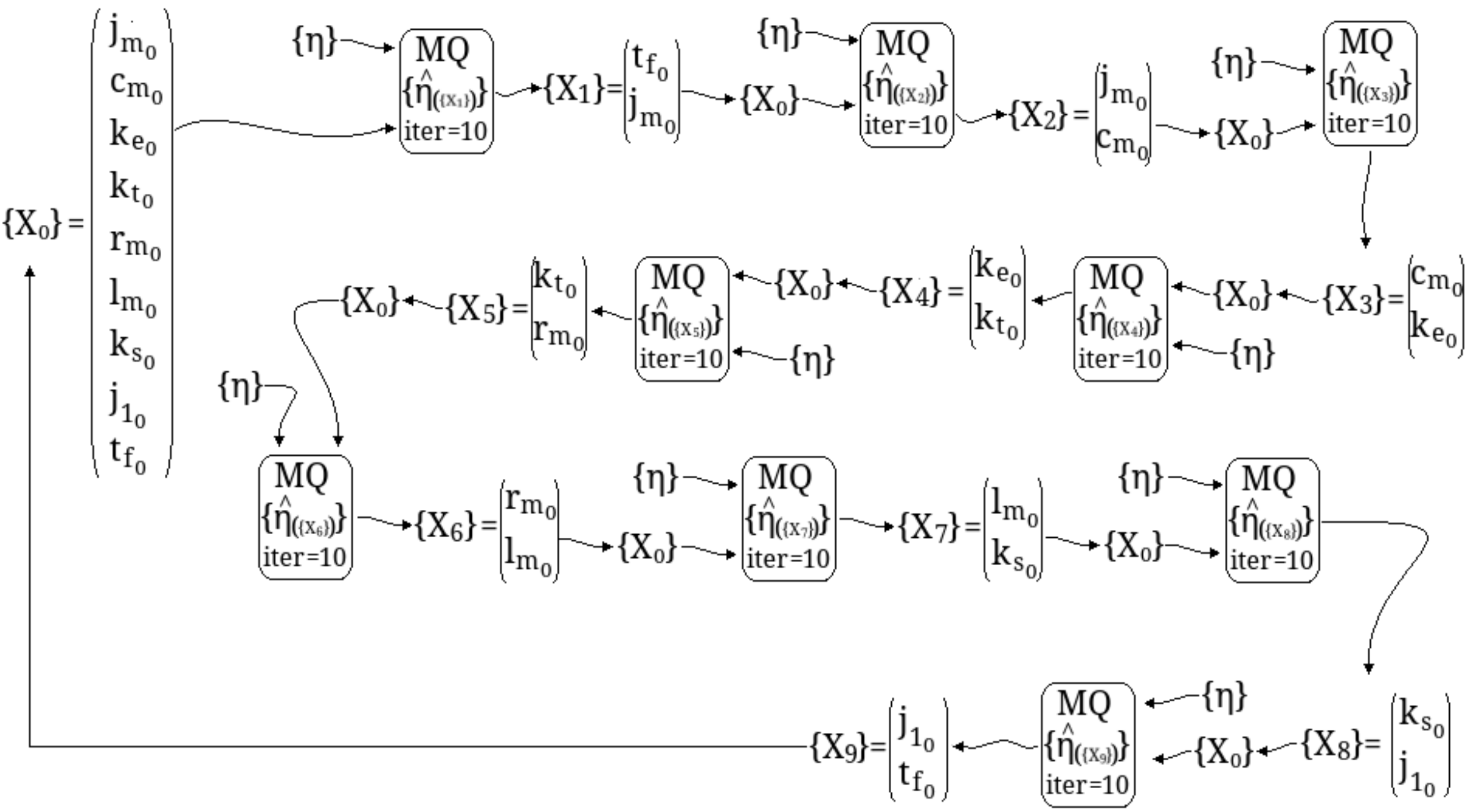}
  \caption{Schematic representation of the proposed heuristic.}
  \label{fig_esq_heristic}
 \end{figure}

The overall procedure is repeated until a steady-state is reached for the misfit function.
A schematic representation of this heuristic can be seen in Figure~\ref{fig_esq_heristic}.
See \cite{Sandoval2013} for further details.

\begin{figure}[h] 
   \centering
   \includegraphics[scale=0.432]{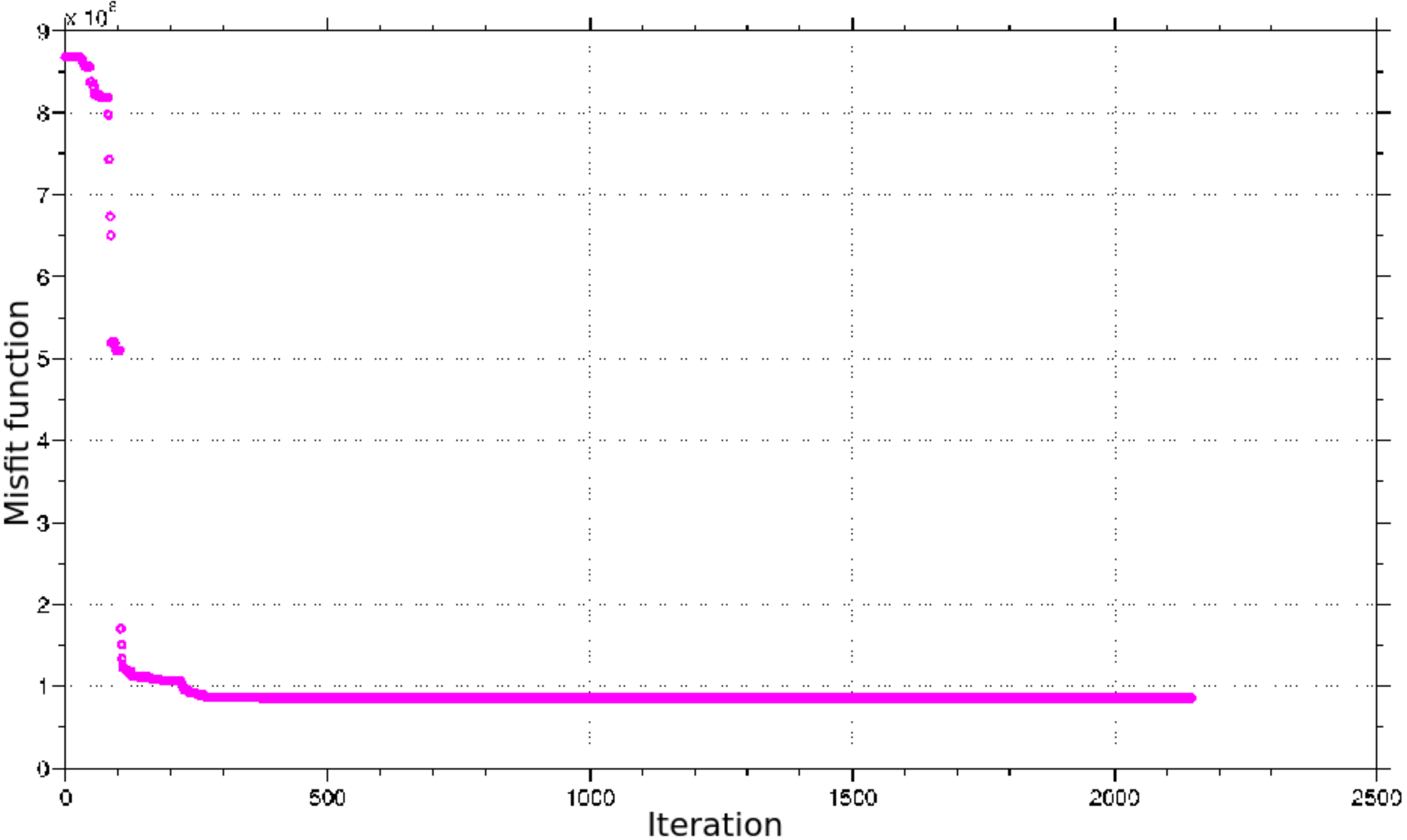}
   \caption{Value of the misfit function for each iteration index of the heuristic algorithm.}
   \label{fig:bncd.MQ.exp.conv}
\end{figure}

\begin{figure}[h] 
  \centering
  \includegraphics[scale=0.484]{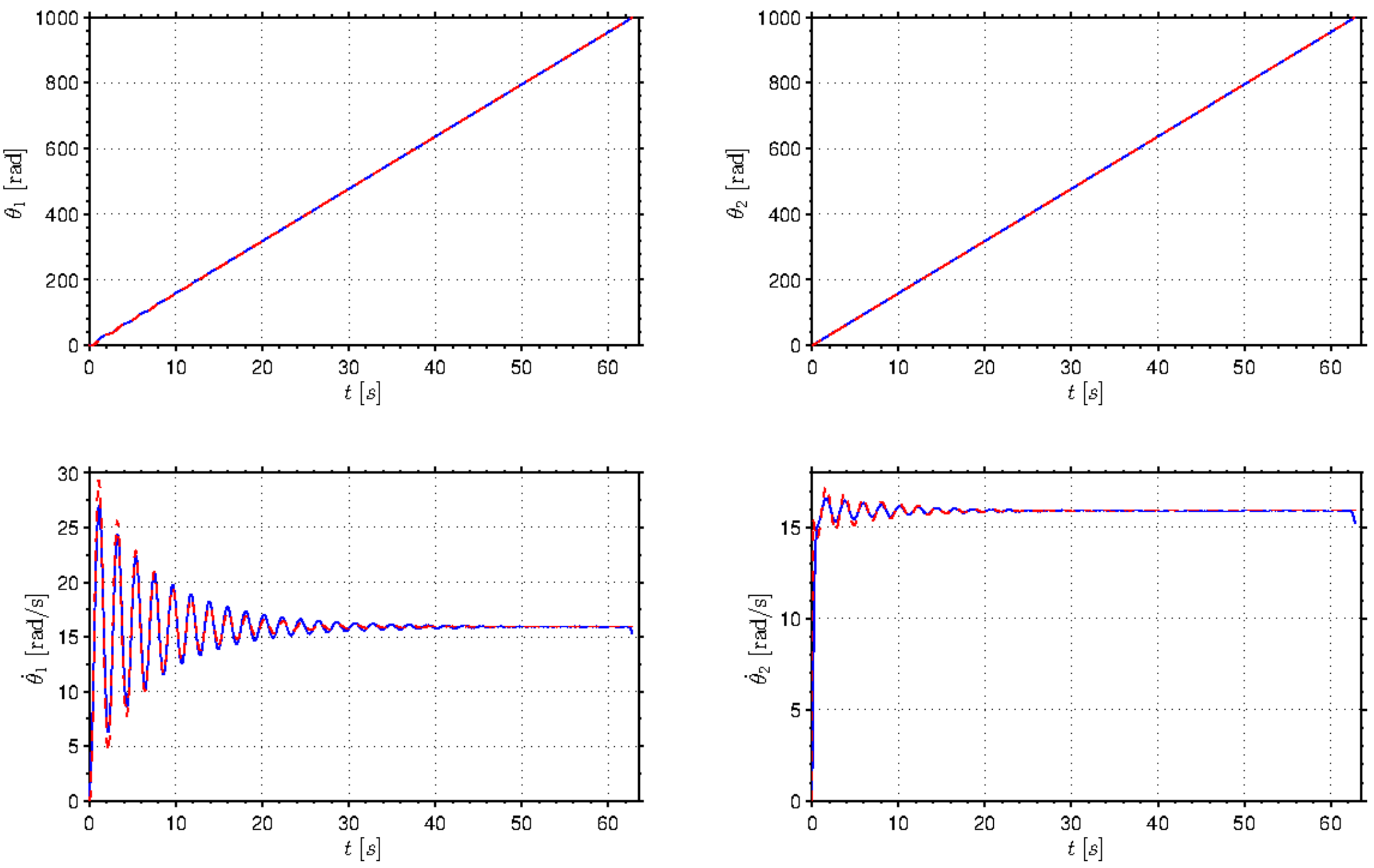} 
  \caption{Values measured (blue line) in the experimental setup and curves (red marks) constructed 
  by the Bayesian estimation method.}
  \label{fig:bncd.MQ.exp}
\end{figure}

\pagebreak
Despite its heuristic nature, the algorithm shows good convergence
properties, as can be seen in the Figures~\ref{fig:bncd.MQ.exp.conv}
and \ref{fig:bncd.MQ.exp}, which respectively show the value of
the misfit function for each iteration, and the comparison between
the rotations and their time derivatives, obtained from the experimental
measurements and from the estimated curves. Note that the good agreement 
between the estimated and measured curves increases the degree of 
confidence in the heuristic estimation.

\pagebreak 
Estimated values for the physical parameters of interest are presented in 
Table~\ref{tab:bancada.est.exp}, as well as the respective initial guess used in the estimation,
and the relative deviations, which are measured with respect to the nominal values shown 
of the Table~\ref{tab_model_parameters}. Note that some parameters exhibit a small relative
deviation, with a magnitude smaller than 7\%, while others estimates are very distant from the 
nominal value, with a deviation of up to 60\%.

Note that the nominal values in the Table~\ref{tab_model_parameters} 
are in fact not the correct value of the parameters of the experimental setup, 
but, in fact, what one believes to know about them. Besides that, as the heuristic 
used showed itself reliable, one can argue that the nominal values of the parameters 
$j_m$, $r_m$, and $l_m$ are not close to the real values of the test rig parameters.

However, it is the authors belief that this discrepancy in deviations has another explanation.
First one can note that all the parameters have large deviations are associated with the motor.
Note also that the modeling of the motor assumes a linear behavior, which does not correspond 
to reality, since torsional dynamics of the shaft is inherently nonlinear due to its coupling with 
the motor. 

Another factor worthy of discussion is the representativeness of a model with 
only two degrees of freedom of rotation to described the dynamics of a flexible shaft, 
which is actually a deformable body with infinite degrees of freedom.

Therefore, it is the main conclusion of this study that the deficiencies of the model 
used to describe the dynamics of the experimental setup are responsible for causing 
the large deviations, with respect to the reference values, observed in the estimated
parameters. The authors believe that better estimates can be obtained if a more realistic 
model is employed. For example, using a system with distributed parameters for modeling 
the shaft and to include a transistor in the description of the motor.

\begin{table}[h]
  \centering
  \begin{tabular}{crrr}
	\toprule
	parameters 	 	&	initial guess &	estimation & relative deviation	\\
	\midrule
	      $j_m$		&	$4.0 \times 10^{-4}$		&	$5.0 \times 10^{-4}$	&	$28.0\,\%$	\\
	      $c_m$		&	$19.0 \times 10^{-5}$	&	$18.0 \times 10^{-5}$	&	$6.6\,\%$	\\
	      $k_e$		&	$601.6 \times 10^{-4}$	&	$587.7 \times 10^{-4}$	&	$2.3\,\%$	\\
	      $k_T$		&	$1.2 \times 10^{-1}$			&	$1.3 \times 10^{-1}$		&	$5.0\,\%$		\\
	      $r_m$		&	$3.3 \times 10^{-1}$			&	$5.3 \times 10^{-1}$			&	$60.0\,\%$	\\
	      $l_m$		&	$1.1 \times 10^{-3}$		&	$0.8 \times 10^{-3}$		&	$23.3\,\%$	\\
	      $k_s$		&	$2.6 \times 10^{-1}$		&	$2.7 \times 10^{-1}$		&	$3.0\,\%$	\\
	      $j_1$		&	$28.3 \times 10^{-3}$	&	$29.8 \times 10^{-3}$		&	$5.4\,\%$	\\
	      $t_f$		&	$1.0 \times 10^{-1}$			&	$1.0 \times 10^{-1}$			&	$0.0\,\%$	\\
\bottomrule
  \end{tabular} 
  \caption{Parameters of the column-motor physical system estimated by the Bayesian estimation method.}
  \label{tab:bancada.est.exp}
 \end{table}

Further results and discussion about the estimation of parameters of the
experimental setup of interest can be seen in \cite{Sandoval2013}.

\section{CONCLUDING REMARKS}
\label{concl_remaks}

In this work was presented a procedure for estimating the parameters 
of an experimental apparatus, composed by a shaft, two rotors, and a
DC motor, that emulates a drilling process. A Bayesian approach was 
used to address the variabilities that are present in the experimental 
measurements due to factors such as measurement error, noise, etc. 

For a verification test, to check the effectiveness of the Bayesian estimation 
technique, first only two parameters were considered as unknown. The results 
obtained with the estimation procedure for this two parameters problem present 
a very low deviation from the nominal values of the experimental apparatus for
both parameters.

On the other hand, in a estimation process which considers nine parameters 
of the experimental apparatus as unknown, convergence problems of the 
Bayesian method were observed. A heuristic procedure was proposed to 
overcome this problem. The estimation results obtained with the proposed 
algorithm show some parameters with low relative deviation with respect 
to the nominal values, and others with large deviation. The presence of these 
large deviations evidentiates a strong deficiency in the model used to describe 
the dynamic behavior of the experimental apparatus, especially by not considering 
the continuous nature of the shaft and nonlinearities intrinsic to the shaft/motor
coupling.

\section*{ACKNOWLEDGMENTS}

The authors are indebted to Brazilian agencies CNPq, CAPES, and FAPERJ 
for the financial support given to this research. They also acknowledge
Mr. Bruno Cayres and Prof. Hans Weber by providing the experimental data 
used in this work.



\end{document}

%% file: macros.tex
\renewcommand{\vec}[1]{\bm{\mbox{#1}}}

\newcommand{\mat}[1]{\left[ #1 \right]}